\documentclass[amsmath,amssymb,aps,nofootinbib,prd,reprint,superscriptaddress]{revtex4-2}

\usepackage[T1]{fontenc}
\usepackage[dvipsnames]{xcolor}
\usepackage{hyperref}
\hypersetup{colorlinks=true,citecolor=RoyalBlue,linkcolor=RoyalBlue,urlcolor=RoyalBlue}

\let\originalleft\left
\let\originalright\right
\renewcommand{\left}{\mathopen{}\mathclose\bgroup\originalleft}
\renewcommand{\right}{\aftergroup\egroup\originalright}
\mathcode`\*="8000
{\catcode`\*=\active\gdef*{\mathclose{}\,\mathopen{}}}

\newcommand{\ab}[1]{\left|#1\right|}
\newcommand{\av}[1]{\left\langle#1\right\rangle}
\newcommand{\br}[1]{\left[#1\right]}
\newcommand{\cu}[1]{\left\{#1\right\}}
\newcommand{\pa}[1]{\left(#1\right)}
\newcommand{\ed}{\mathop{}\!\mathrm{d}}
\newcommand{\pd}{\mathop{}\!\partial}
\newcommand{\sg}{\operatorname{sg}}
\newcommand{\tlam}{\tilde{\lambda}}
\newcommand{\TT}{\mathrm{T}}
\newcommand{\hTT}{\widehat{\TT}}
\newcommand{\Rnn}{\mathcal{R}_{n,n-1}}
\newcommand{\Trees}{\mathsf{Trees}}
\newcommand{\IncTrees}{\mathsf{IncTrees}}

\begin{document}

\title{Single-minus graviton tree amplitudes are nonzero}

\author{
Alfredo Guevara,\textsuperscript{1}
Alexandru Lupsasca,\textsuperscript{2,3}
David Skinner,\textsuperscript{4}\\
Andrew Strominger,\textsuperscript{5}
and Kevin Weil\textsuperscript{2}
on behalf of OpenAI\\
\scriptsize
\textsuperscript{1}Institute for Advanced Study\quad
\textsuperscript{2}OpenAI\quad
\textsuperscript{3}Vanderbilt University\quad
\textsuperscript{4}Cambridge University\quad
\textsuperscript{5}Harvard University}
\noaffiliation

\begin{abstract}
Single-minus tree-level $n$-graviton scattering amplitudes are revisited.
Often presumed to vanish, they are shown here to be nonvanishing for certain ``half-collinear'' configurations existing in Klein space or for complexified momenta.
A Berends--Giele recursion relation for these amplitudes is derived and solved in a form involving a sum over trees.
In a restricted kinematic decay region, this solution simplifies significantly to an $(n{-}2)$-fold product of soft factors.
It is further shown in this region that, combined with suitable analyticity assumptions, the $n$-graviton amplitude is generated by a recursive $\mathcal{L}w_{1+\infty}$ Ward identity with the three-graviton amplitude as a seed.
\end{abstract}

\maketitle

Reconciling Einstein gravity with quantum mechanics is a central challenge in modern physics.
\emph{Self-dual} gravity~\cite{MasonWoodhouse,Krasnov:2016} provides a much more manageable---while still rich---toy model for addressing this challenge. Quantum effects are finite, computable, and one-loop exact~\cite{Bern:1998sv,Chalmers:1996rq,Doran:2023anomaly}.
A complete solution of quantum self-dual gravity is conceivably within reach~\cite{Ooguri:1990ww,Ball:2021tmb,Costello:2022wso,Bittleston:2022nfr,Bittleston:2022jeq,Fernandez:2024qnu} and may shed light on quantum Einstein gravity.  

Penrose famously solved classical self-dual gravity using twistor theory \cite{Penrose:1976,Penrose:1976js} a half-century ago.
The highly non-trivial solutions are generated by an infinite-dimensional symmetry group known as $\mathcal{L}w_{1+\infty}$ \cite{Adamo:2021lrv}, which also appears in Einstein gravity   \cite{Guevara:2021abz,Strominger:2021mtt}.

It is sometimes stated that the tree amplitudes of self-dual gravity are nonzero only for three or fewer gravitons \cite{Chalmers:1996rq}.
Tree amplitudes are purportedly a repackaging of the classical solutions.
This raises a conundrum: how can the richness of the nonlinear Penrose solutions possibly be encoded in the nearly trivial tree scattering amplitudes?

In this paper, we show that the self-dual tree-level amplitudes are in fact nonzero for any number of gravitons.
They are supported on single-minus half-collinear configurations.
This means that they have exactly one minus-helicity graviton and in split signature are localized to a null line on the boundary of spacetime. A general formula is derived for the single-minus amplitude involving sums over tree diagrams with a number of terms that grows exponentially in the number $n$ of gravitons.
In a kinematically restricted ``decay region'' we find a simple formula in terms of a product of soft factors.
These nonzero single-minus amplitudes extend to gravity a similar result for Yang--Mills theory \cite{Guevara:2026qzd}. 

The symmetry group $\mathcal{L}w_{1+\infty}$ plays a central role.
The $\mathcal{L}w_{1+\infty}$ Ward identities are a tower of soft theorems that recursively relate the $n{+}1$ to the
$n$-graviton scattering amplitude.
This $\mathcal{L}w_{1+\infty}$ recursion, along with some analyticity assumptions, enables a complete construction of the decay-region\footnote{We do not know if this is possible outside of the decay region.} self-dual  amplitudes from a three-graviton seed.
This beautifully mirrors the Penrose construction of the self-dual solutions from the action of $\mathcal{L}w_{1+\infty}$ on the vacuum. 

The classical solutions and tree amplitudes of self-dual gravity are a small subset of those of (complexified) Einstein gravity.
Both theories admit an $\mathcal{L}w_{1+\infty}$ action.
In the Einstein case, $\mathcal{L}w_{1+\infty}$ was recently shown \cite{Guevara:2025tsm} to recursively generate all the double-minus amplitudes. The current work extends this statement to single-minus and sheds further light on the incompletely understood role of $\mathcal{L}w_{1+\infty}$ in Einstein gravity. 

The workhorse of this paper is an adaptation of the Berends--Giele recursion relation~\cite{Berends:1987me}, a rewriting of the Feynman rules, to the single-minus context---see~\eqref{eq:Recursion}.
Using this recursion, the expression for the general single-minus amplitude is derived and presented in~\eqref{eq:Full}.
Restricting to a special kinematic decay region (see Sec.~\ref{sec:Bootstrap}), in which one particle is ingoing and all others outgoing, we obtain the much simpler expression
\begin{align}
    \label{eq:Beauty}
    \mathcal{M}_n=i^{2-n}\prod_{a=1}^{n-2}\sum_{[aj]>0}[aj]\,\prod_{b=1}^{n-1}\delta\pa{\av{bn}}\,\delta^2\!\pa{\sum_{i=1}^n\tlam_i}.
\end{align}
Here, the $n^\text{th}$ graviton is minus-helicity, and the rest of our notation is explained below. This formula is analytic (apart from the $\delta$-functions) in the noncollinear momenta except for the fully collinear locus ($[ij]=0$), where the first derivative has a $\delta$-function.  Simplification of the general solution \eqref{eq:Full}, if possible,  is left to future work. 

After a brief review of our notation, Sec.~\ref{sec:Amplitudes} gives the general form of half-collinear amplitudes and defines a fully permutation-invariant stripped amplitude $M_{1\cdots n}$ containing the essential information.
The decay region is defined in Sec.~\ref{sec:Bootstrap} and the formula \eqref{eq:Beauty} for $\mathcal{M}_n$ is derived therein from $\mathcal{L}w_{1+\infty}$ recursion.
Next, the general-region Berends--Giele recursion relation is derived in Sec.~\ref{sec:Recursion}, and Sec.~\ref{sec:Formula} solves it in the decay region.
Lastly, App.~\ref{app:MHV} reviews MHV amplitudes and Cayley trees, while App.~\ref{app:Derivation} and App.~\ref{app:ward-recursion-proof} supply details of the Berends--Giele and $\mathcal{L}w_{1+\infty}$ recursion formulae, respectively.

Both GPT-5.2 Pro and a new OpenAI internal model played a significant role at all stages of this project.

\paragraph*{Notation and useful identities.}

We use spinor-helicity variables for massless momenta, with $(\lambda,\tlam)$ denoting \emph{real} spinors in $(2,2)$ signature,
\begin{align}
    p_{\alpha\dot\alpha}=\lambda_\alpha\tlam_{\dot\alpha}.
\end{align}
We fix a Lorentz and little-group frame by setting
\begin{align}
    \label{eq:Conventions}
    |i\rangle=\lambda_i=\pa{1,z_i},\qquad
    |i]=\tlam_i=\omega_i\pa{1,\tilde{z}_i},
\end{align}
with $z_i$ and $\tilde z_i$ real and independent, and $\omega_i\in\mathbb{R}$.
We use the standard brackets
\begin{align}
    \langle ij\rangle&=\av{\lambda_i\lambda_j}
    =\epsilon_{\alpha\beta}\lambda_i^\alpha\lambda_j^\beta,\\
    [ij]&=\br{\tlam_i\tlam_j}
    =\epsilon_{\dot\alpha\dot\beta}\tlam_i^{\dot\alpha}\tlam_j^{\dot\beta},
\end{align}
such that
\begin{align}
    \label{eq:Momentum}
    p_{ij}^2=(p_i+p_j)^2=\langle ij\rangle[ij].
\end{align}
With respect to the parameterization~\eqref{eq:Conventions}, we have
\begin{align}
    \av{ij}=z_{ij},\qquad
    [ij]=\omega_i\omega_j\tilde{z}_{ij},
\end{align}
where $z_{ij}:=z_i-z_j$ and $\tilde{z}_{ij}:=\tilde{z}_i-\tilde{z}_j$.

We normalize all $\delta$-functions by
\begin{align}
    \label{eq:DeltaNormalization}
    \int\delta(x)\ed x=2\pi,\qquad
    \frac{1}{x+i\epsilon}-\frac{1}{x-i\epsilon}\stackrel{\epsilon\to0}{=}-i\,\delta(x),
\end{align}
and use the standard Feynman prescription $1/(p^2+i0)$.

For later use, it is convenient to define, for any finite set $S\subset\cu{1,\ldots,n}$ with $|S|\ge 1$,
\begin{align}
    \label{eq:Notation}
    \tlam_S:=\sum_{i\in S}\tlam_i,\qquad
    [S,T]:=\br{\tlam_S\tlam_T}.
\end{align}

\section{Single-minus graviton amplitudes}
\label{sec:Amplitudes}

We now consider the single-minus graviton amplitude with one minus-helicity leg at label $n$,
\begin{align}
    \mathcal{M}_n:=\mathcal{M}_n(1^+,\ldots,(n{-}1)^+,n^-).
\end{align}
The \emph{half-collinear regime} is defined by
\begin{align}
    \label{eq:HalfCollinear}
    \av{ij}=0\qquad\forall\,i,j\in\cu{1,\ldots,n}.
\end{align}
In Klein signature, this is compatible with nonzero $[ij]$.
In the frame~\eqref{eq:Conventions}, the condition~\eqref{eq:HalfCollinear} forces $z_{ij}=0$ but does not restrict either $\omega_i$ nor $\tilde{z}_i$.

The fact that the amplitudes can only be supported on this locus follows the same argument as in \cite{Guevara:2026qzd}.
In short, the usual power-counting argument for the vanishing of single-minus tree amplitudes relies on choosing a reference spinor that renders polarization vectors mutually orthogonal, but there are insufficient powers of momenta in the numerator to be able to saturate all contractions.
In the half-collinear regime, the choice is obstructed because $\langle jn\rangle=0$ for $j\neq n$ makes the plus-helicity polarizations singular.
As in Yang--Mills theory, this loophole allows nontrivial distributional support on the locus~\eqref{eq:HalfCollinear}.

In the half-collinear regime, it is convenient to factor out from $\mathcal{M}_n$ its universal dependence on the collinear variables and hence define a \emph{stripped amplitude} $M_{1\cdots n}$ by
\begin{align}
    \label{eq:AnsatzP}
    \mathcal{M}_n&=i^{2-n}\frac{\av{rn}^{n+3}}{\av{r1}^3\av{r2}^3\cdots\av{r\,n{-}1}^3}M_{1\cdots n}\notag \\
    &\qquad\times\prod_{a=1}^{n-1}\delta\pa{\av{an}}\,\delta^2\!\pa{\sum_{i=1}^n\av{ri}\tlam_i}.
\end{align}
The prefactor carries the required little-group scaling for one minus and $n-1$ plus gravitons, so that $M_{1\cdots n}$ has no little-group weight in the external particles.\footnote{With the normalization~\eqref{eq:AnsatzP}, $M_{1\cdots n}$ scales with weight $n-1$ in the reference spinor $|r\rangle$.}
On the support of the collinear $\delta$-functions, this expression is independent of $|r\rangle$ provided $\av{rn}\neq 0$.
In the fixed frame~\eqref{eq:Conventions}, if $|r\rangle=(0,1)$, then $\av{ri}=1$ for all $i$,  and so~\eqref{eq:AnsatzP} becomes
\begin{align}
    \label{eq:Ansatz}
    \mathcal{M}_n=i^{2-n}M_{1\cdots n}\prod_{a=1}^{n-1}\delta(z_{an})\,\delta^2\!\pa{\sum_{i=1}^n\tlam_i}.
\end{align}
We will study $M_{1\cdots n}$, which depends only on $\{\tlam_i\}$.
We also define a set-collinear distribution
\begin{align}
    \label{eq:DeltaS}
    \delta_S:=i^{1-|S|}\prod_{i\in S\setminus\cu{r(S)}}\delta\pa{z_i-z_{r(S)}},
\end{align}
where $r(S)\in S$ is an arbitrary chosen ``root'' label.
On the support of $\delta_S$, all $|i\rangle$ for $i\in S$ coincide, so $\tlam_S$ in \eqref{eq:Notation} is the only nontrivial spinor data in the block $S$.

\section{\texorpdfstring{$\mathcal{L}w_{1+\infty}$}{Lw(1+∞)} bootstrap}
\label{sec:Bootstrap}

Three-point amplitudes are famously fixed by Poincar\'e invariance.
In this section, we ask if single-minus amplitudes could perhaps be fixed by a larger symmetry.
A sharp candidate for this is the recently identified $\mathcal{L}w_{1+\infty}$ symmetry of gravity \cite{Guevara:2021abz,Strominger:2021mtt}.
We find the answer to be positive.
More precisely, we posit two assumptions detailed in the following two subsections, and use them to determine a closed-form expression immediately afterwards.

\subsection{Chamber analyticity and the decay region}

As a function of the coordinates $\tilde{z}_k$, the (stripped) amplitude $M_{1\cdots n}$ is analytic everywhere, except possibly at
\begin{itemize} 
    \item $[ij]=0$, in particular the coincidence loci $\tilde{z}_i=\tilde{z}_j$,
    \item ``bulk'' loci where weighted sums coincide, namely the vanishing of \eqref{eq:Notation}, or in particular,
\begin{align}
    \label{eq:Locus}
    \sum_{k\in S}\frac{\omega_k}{\omega_S}\tilde{z}_k=\sum_{k\in T}\frac{\omega_k}{\omega_T}\tilde{z}_k.
\end{align}
\end{itemize}
On such loci, $M_{1\cdots n}$ may be singular or discontinuous.

To avoid such singularities, we introduce a restricted kinematic region $\Rnn^\pi$ as follows.
As in~\cite{Guevara:2026qzd}, we first define a kinematic region by demanding that there exists a Lorentz frame in which
\begin{align}
    \label{eq:RnnOmega}
    \omega_n<0,\qquad
    \omega_a>0,\qquad
    a\in\cu{1,\ldots,n{-}1}.
\end{align}
By momentum conservation $\sum_{i=1}^n\tlam_i=0$, we can write
\begin{align}
    \label{eq:Average}
    \omega_n=-\sum_{a=1}^{n-1}\omega_a,\qquad
    \tilde{z}_n=\frac{\sum_{a=1}^{n-1}\omega_a\tilde{z}_a}{\sum_{a=1}^{n-1}\omega_a},
\end{align}
so that $\tilde{z}_n$ is the $\omega$-weighted average of the outgoing $\tilde{z}_a$.
For gravity, we now define a further subregion $\Rnn^\pi$ of \eqref{eq:RnnOmega} by additionally requiring a unique outlier among the outgoing $\tilde{z}$'s, which we take to be $\tilde{z}_{n-1}$.
Then we order the remaining $\tilde{z}$'s according to a fixed permutation $\pi$ of $\{1,2,\ldots,n-2\}$.
That is, we have
\begin{align}
    \label{eq:Outlier}
    \tilde{z}_{\pi(1)}<\tilde{z}_{\pi(2)}<\cdots<\tilde{z}_{\pi(n-2)}<\tilde{z}_n<\tilde{z}_{n-1}.
\end{align}
In this region $\Rnn^\pi$, all $[ij]$ have fixed signs and hence are locally linear.
A remarkable property, shown explicitly in Sec.~\ref{sec:Formula} below, is that even though $\Rnn^\pi$ contains bulk loci \eqref{eq:Locus}, $M_{1\cdots n}$ is not singular within this region, which is thus a genuine chamber.
We will work inside this chamber in the rest of this section.

It is also useful to introduce the permutation-invariant decay region
\begin{align}
    &\mathcal{D}_{n,n-1}=\bigcup_{\pi\in S_{n-2}}\Rnn^\pi\\
    &\quad=\Big\{\omega_n<0,\ \omega_a>0,\quad\forall\,a\in\cu{1,\dots,n-1},\notag\\
    &\qquad\quad\tilde{z}_j<\tilde{z}_n<\tilde{z}_{n-1},\quad\forall\,j\in\cu{1,\dots,n-2}\Big\}.
\end{align}
This region contains internal walls where $M_{1\cdots n}$ may fail to be analytic, but one can provide a general formula for the amplitude by permuting the labels of its internal components $\Rnn^\pi$.
For the remainder of this paper, we assume $\pi=(1,\ldots,n{-}2)$ without loss of generality and denote $\Rnn^\pi$ by the shorthand $\Rnn$.

\subsection{\texorpdfstring{$\mathcal{L}w_{1+\infty}$}{Lw(1+∞)} Ward identities}

The presence of $\mathcal{L}w_{1+\infty}$ symmetry, which acts through Ward identities, imposes strong constraints on gravitational scattering.
In fact, these constraints have recently been found to fully determine amplitudes in momentum space~\cite{Guevara:2025tsm,Guevara:2022qnm}.
In short, for the particular case of (stripped) MHV amplitudes, the $n$-point formula is obtained from that with $n-1$ points by the insertion of an all-orders-soft graviton with label $s$.
Formally, this reads~\cite{Guevara:2025tsm}
\begin{align}
    \label{eq:Ward}
    \tilde{M}_{1\cdots n}=\sum_{i\neq s}^{n}\frac{[si]^2}{p_{si}^2+i\epsilon}\tilde{M}_{\widehat{s}}^{(i)},\quad
    \tilde{M}_{\widehat{s}}^{(i)}:=\tilde{M}_{\widehat{s}}\Big|_{\tlam_i\to\tlam_i+\tlam_s}.
\end{align}
Here, as in App.~\ref{app:MHV}, $\tilde{M}_{1\cdots n}$ denotes the amplitude stripped of momentum conservation, while the shorthand $\tilde{M}_{\widehat{s}}$ stands for an amplitude with $n{-}1$ labels (\emph{i.e.}, no leg $s$).
We refer to \eqref{eq:Ward} as the $\mathcal{L}w_{1+\infty}$ Ward identities.

Our assumption is that the same identities \eqref{eq:Ward} can be applied to the single-minus amplitudes.
In this case, the object of interest is the amplitude \eqref{eq:Ansatz} stripped of the half-collinear and half-momentum-conserving $\delta$-function.
It follows from the identities \eqref{eq:Momentum}--\eqref{eq:DeltaNormalization} that, inside a fixed chamber, the above expression reduces to
\begin{align}
    \label{eq:w}
    M_{1\cdots n}=\frac{1}{2}\sum_{\substack{i=1\\i\neq s}}^{n}|[si]|M_{\widehat{s}}^{(i)},\quad
    M_{\widehat{s}}^{(i)}:=M_{\widehat{s}}\Big|_{\tlam_i\to \tlam_i+\tlam_s},
\end{align}
where, as usual, $|x|$ denotes the absolute value of $x$.
The Ward identity is to be interpreted in terms of the tower of differential operators $\omega_s^k\sim\pa{\tlam_s\frac{\pd}{\pd\tlam_i}}^k$ generating the shift, namely the $\mathcal{L}w_{1+\infty}$ generators.\footnote{At the exact location of singularities, these operators generate distributional terms at the wall, which we ignore in the following.}
In the following, we will extend this relation inside a fixed chamber assuming that the shifts $\tlam_i\to\tlam_i+\tlam_s$ stay within their walls, \emph{e.g.}, $\omega_s\ll\omega_i$, so the Ward identity is unambiguous. 

\subsection{Decay region amplitude}
 
With the Poincar\'e-fixed seed $M_{123}=|[12]|$, define for $a\in\cu{1,\dots,n{-}2}$ the soft factors
\begin{align}
    S_a:=\frac12\sum_{j=1}^n|[aj]|.
\end{align}
Then the Ward recursion generates
\begin{align}
    M_{1\cdots n}\Big|_{\mathcal{D}_{n,n-1}}=\prod_{a=1}^{n-2}S_a.
\end{align}
This formula is the gravitational analogue of the inverse soft construction found in \cite{Guevara:2026qzd}.
It exhibits $S_{n-2}$ permutation invariance inherited from $\mathcal{D}_{n,n-1}$.
It is proven by deriving the corresponding expression in the chamber $\Rnn$ and then using permutation symmetry to extend the result to arbitrary $\pi$.
Note that in that case, we have
\begin{align}
    S_a\Big|_{\Rnn}=\sum_{j=a+1}^{n-1}[ja],
\end{align}
resulting in
\begin{align}
    \label{eq:ProductFormula}
    M_{1\cdots n}\Big|_{\Rnn}=\prod_{i=1}^{n-2}\pa{\sum_{j=i+1}^{n-1}[ji]}.
\end{align}
The idea of the proof is conceptually simple.
Take any label $\tlam_s\to0$, landing on the analogous chamber $\Rnn^{(s)}$ without label $s$.
We proceed by induction and assume 
\begin{align}
    \label{eq:Induction}
    M_{\hat{s}}\Big|_{\Rnn^{(s)}}=\prod_{i=1,i\neq s}^{n-2}\pa{\sum_{j=i+1,j\neq s}^{n-1}[ji]}.
\end{align}
In App.~\ref{app:ward-recursion-proof}, we apply the Ward identity \eqref{eq:w} and derive the result \eqref{eq:ProductFormula}.
This assumes $\omega_s\ll\omega_a$ so that we do not cross any wall.
We then use our assumption that the amplitude is analytic in $\Rnn$, which has no walls, to extend the result throughout that chamber.

This concludes the first main result of this paper.
In the following, we will derive a more general formula that holds without any restriction on the external kinematics (beyond momentum conservation and being in the half-collinear regime).
We obtain this more general result by an explicit calculation akin to summing all contributing Feynman diagrams.
It would be very interesting to understand if we can relax our analyticity assumptions and use $\mathcal{L}w_{1+\infty}$ to fix $M_{1\cdots n}$ with general kinematics.

\section{Berends--Giele recursion}
\label{sec:Recursion}

The single-minus graviton sector is governed by an unordered Berends--Giele recursion for gravitational form factors with one leg off-shell.
Starting from this, we derive in App.~\ref{app:Derivation} a fully \emph{on-shell} recursion for $M_{1\cdots n}$.
This on-shell recursion is equivalent to summing tree-level Feynman diagrams, but is tailored to the half-collinear regime and is analogous to the construction of \cite{Guevara:2026qzd} in the case of Yang--Mills gauge theory.

To state the result, we first define for nonempty sets $S$ a family of \emph{preamplitudes} $\bar{M}_S$ by
\begin{align}
    \bar{M}_{\cu{i}}=1,\qquad
    \bar{M}_{\cu{i,j}}=0,
\end{align}
and, for $|S|\ge 3$,
\begin{align}
    \label{eq:Preamplitudes}
    \bar{M}_S=-\sum_{\substack{S=S_1\sqcup\cdots\sqcup S_A\\ A\ge 3}}V_{\tlam_{S_1}\cdots\tlam_{S_A}}\prod_{a=1}^{A}\bar{M}_{S_a},
\end{align}
where the sum is over all set partitions into $A\ge 3$ nonempty blocks.
Here, $V_{\tlam_{S_1}\cdots\tlam_{S_A}}$ is a ``retarded'' multi-point vertex, which we will now define in terms of a sum over Cayley (\emph{i.e.}, spanning) trees.
In App.~\ref{app:MHV}, we review the role of similar spanning trees in the familiar MHV amplitude for gravity~\cite{Nguyen:2009jk,Hodges:2011wm,Feng:2012sy}.

Let $\Trees(A)$ denote the set of Cayley trees on the vertex set $\cu{1,\ldots,A}$.
For a tree $T\in\Trees(A)$ and an edge $e=(u,v)\in E(T)$, removing $e$ disconnects the vertex set into two components $A_e\sqcup B_e=\cu{1,\ldots,A}$; we take $A_e$ to be the component with the smaller label among $\cu{u,v}$, which we take to be $u$.

We define
\begin{align}
    \label{eq:VertexDefinition}
    V_{\tlam_1\cdots \tlam_A}:=\sum_{T\in\Trees(A)}\,\prod_{e=(u,v)\in E(T)}\ab{\br{uv}}\,\Theta\pa{-\frac{\br{A_eB_e}}{\br{uv}}},
\end{align}
with $V_{\tlam_1}:=1$ and $\Theta(x)$ denoting the step function as usual.
Similarly, we define the ``advanced'' vertex $\bar{V}$ by flipping the sign in the $\Theta$-argument,
\begin{align}
    \label{eq:VertexBarDefinition}
    \bar{V}_{\tlam_1\cdots\tlam_A}:=\sum_{T\in\Trees(A)}\,\prod_{e=(u,v)\in E(T)}\ab{\br{uv}}\,\Theta\pa{+\frac{\br{A_eB_e}}{\br{uv}}},
\end{align}
with $\bar{V}_{\tlam_1}:=1$.
Finally, we introduce
\begin{align}
    \label{eq:HatT}
    \hTT_{\tlam_1\cdots\tlam_A}:=V_{\tlam_1\cdots\tlam_A}-\bar{V}_{\tlam_1\cdots\tlam_A}.
\end{align}
As explained in App.~\ref{app:Derivation}, this is the on-shell kernel that appears after LSZ reduction of BG recursion.

Having defined $\bar{M}_S$, the stripped amplitude $M_{1\cdots n}$ is given by the recursion
\begin{align}
    \label{eq:Recursion}
    M_{1\cdots n}=-\sum_{\substack{\cu{1,\ldots,n-1}=S_1\sqcup\cdots\sqcup S_A\\ A\ge 2}}\!\hTT_{\tlam_{S_1}\cdots\tlam_{S_A}}\,\prod_{a=1}^{A}\bar{M}_{S_a}.
\end{align}
This on-shell recursion relation is the next main result of the paper.
When combined with the prefactor~\eqref{eq:Ansatz}, it provides a general, albeit implicit, formula for the $n$-particle single-minus gravity amplitude with arbitrary external kinematics.
It is the gravitational analogue of the ordered single-minus recursion in gauge theory~\cite{Guevara:2026qzd}, with color-ordering replaced by set partitions and Parke--Taylor factors replaced by $\hTT$.
An explicit non-recursive form is also provided in App.~\ref{app:Derivation}.

In a general kinematic region, it is worth noting that, because the helicity dependence has been absorbed in our frame~\eqref{eq:Ansatz}, the stripped amplitude $M_{1\cdots n}$ is fully permutation-invariant, 
\begin{align}
    M_{\sigma(1)\cdots\sigma(n)}=M_{1\cdots n},\qquad \forall\,\sigma\in S_{n}.
\end{align}
This property holds but is not manifest in the LSZ formula \eqref{eq:Recursion}.

\subsection{Concrete examples}

From~\eqref{eq:Recursion}, the lowest-point stripped amplitudes are as follows.
At three points,
\begin{align}
    M_{12 3}=\ab{[12]}.
\end{align}
At four points, since $\bar{M}_{\cu{i,j}}=0$, only the all-singleton partition contributes,
\begin{align}
    M_{1234}= -\,\hTT_{\tlam_1\tlam_2\tlam_3}
    =-\Big(V_{\tlam_1\tlam_2\tlam_3}-\bar{V}_{\tlam_1\tlam_2\tlam_3}\Big),
\end{align}
where each three-vertex is a sum over the three trees on $\cu{1,2,3}$.
By some algebraic manipulation of the step functions, one may turn this into
\begin{align}
    M_{1234}=\frac{1}{2}\Big(|[12]|\,|[34]|+|[13]|\,|[24]|+|[14]|\,|[23]|\Big).
\end{align}
At five points,
\begin{align}
    M_{12345}=-\hTT_{1234}
    &-|[15]|V_{234}-|[25]|V_{134}\notag\\
    &-|[35]|V_{124}-|[45]|V_{123}.
\end{align}
Here, expanding the vertices fully explicitly generates 44 terms.
These expressions rapidly proliferate in the generic region.
Nevertheless, we find that, when specialized to the region $\mathcal{R}_{5,4}$, this five-particle amplitude is just
\begin{align}
    M_{12345}\Big|_{\mathcal{R}_{5,4}}\!=\!\Big([21]+[31]+[41]\Big)\Big([32]+[42]\Big)[43].
\end{align}
in agreement with \eqref{eq:Induction}. 

\section{Amplitudes in the chamber}
\label{sec:Formula}

In this section, we elaborate on the chamber $\Rnn$ in which the recursion~\eqref{eq:Recursion} collapses and the answer admits a closed form.
This rederives~\eqref{eq:ProductFormula} by merging the three steps analogous to our gauge-theory analysis \cite{Guevara:2026qzd}: (i) vanishing of $V$ on outgoing data, (ii) collapse of the recursion, and (iii) evaluation of $\bar{V}$.
A new ingredient that emerges in this last step is the application of the directed matrix-tree theorem.

\subsection{Vanishing of \texorpdfstring{$V$}{V} in the chamber \texorpdfstring{$\Rnn$}{R(n,n-1)}}

We first show that within $\Rnn$, the retarded vertices built only from outgoing data vanish:
\begin{align}
    \label{eq:VanishSubset}
    V_{\tlam_S}\Big|_{\Rnn}=0,
\end{align}
for any set $S\subseteq\cu{1,\ldots,n-1}$ with $|S|\ge 2$.
This eliminates most walls of the type \eqref{eq:Locus} and in conjunction with the next subsection shows that this is a genuine chamber.

The proof closely follows \cite{Guevara:2026qzd}.
Write $\tlam_i=\omega_i(1,\tilde{z}_i)$ with $\omega_i>0$ for $i\in S$.
For a cut $A\sqcup B=S$, define weighted means $\tilde{z}_A=\frac{1}{\omega_A}\sum_{i\in A}\omega_i\tilde{z}_i$, so that
\begin{align}
    [A,B]=\omega_A\omega_B(\tilde{z}_A-\tilde{z}_B),\quad
    [uv]=\omega_u\omega_v(\tilde{z}_u-\tilde{z}_v).
\end{align}
For each fixed tree $T$ appearing in~\eqref{eq:VertexDefinition}, at least one edge $e=(u,v)$ obeys
\begin{align}
    (\tilde{z}_{A_e}-\tilde{z}_{B_e})(\tilde{z}_u-\tilde{z}_v)>0,
\end{align}
as follows from a tree weighted-variance identity.
For that edge, the ratio $\frac{[A_e,B_e]}{[uv]}$ is positive and the corresponding factor $\Theta\pa{-\frac{[A_e,B_e]}{[uv]}}$ vanishes, so that $V_T=0$, and hence~\eqref{eq:VanishSubset} holds.

It follows from~\eqref{eq:Preamplitudes} that in the chamber $\Rnn$,
\begin{align}
    \bar{M}_S\Big|_{\Rnn}=0\qquad\text{for all }|S|\ge 2,
\end{align}
while $\bar{M}_{\cu{i}}=1$ remains.
Therefore, in~\eqref{eq:Recursion} only the all-singleton partition contributes, and as such we obtain
\begin{align}
    \label{eq:Collapse}
    M_{1\cdots n}\Big|_{\Rnn}=-\hTT_{\tlam_1\cdots\tlam_{n-1}}\Big|_{\Rnn}
    =\bar{V}_{\tlam_1\cdots\tlam_{n-1}}\Big|_{\Rnn}.
\end{align}

\subsection{Evaluating \texorpdfstring{$\bar{V}$}{V} in the chamber \texorpdfstring{$\Rnn$}{R(n,n-1)}}

We can now simplify the vertex \eqref{eq:VertexBarDefinition} in $\Rnn$.
Using momentum conservation $\tlam_{B_e}=-\tlam_n-\tlam_{A_e}$, we rewrite
\begin{align}
    [A_e,B_e]=[\tlam_n,\tlam_{A_e}]
    =[n,A_e],
\end{align}
so that plugging \eqref{eq:VertexBarDefinition} into \eqref{eq:Collapse}, we get\footnote{We emphasize that we cannot yet apply the matrix-tree theorem to this formula because each factor in the product depends on the global structure of the tree.
This precludes the existence of a simple Hodges formula and is the motivation for introducing the outlier condition under which the step functions become unity.}
\begin{align}
    \label{eq:VbarRn}
    M_{1\cdots n}\Big|_{\Rnn}
    \!\!=\!\!\sum_{T\in\Trees(n-1)}\,\prod_{e=(u,v)\in E(T)}
    \ab{[uv]}\Theta\pa{\frac{[n,A_e]}{[uv]}}.
\end{align}
Next, we use the outlier chamber condition~\eqref{eq:Outlier}.
One can convince oneself that in $\Rnn$, the sign of $[n,A]$ depends only on whether $n-1\in A$:
\begin{align}
    \sg([n,A])=
    \begin{cases}
    -1,& n-1\notin A,\\
    +1,& n-1\in A.
    \end{cases}
\end{align}
With labels ordered as in~\eqref{eq:Outlier}, we have $[uv]<0$ for all $u<v\le n-1$.
Hence, each step function in~\eqref{eq:VbarRn} is just
\begin{align}
    \Theta\pa{\frac{[n,A_e]}{[uv]}}=1
    \;\Leftrightarrow\;
    [n,A_e]<0
    \;\Leftrightarrow\;
    n{-}1\notin A_e.
\end{align}
Thus, only those trees survive for which, for every edge $e=(u,v)$ with $u<v$, the root $n-1$ lies in the component containing $v$.
Equivalently, the contributing trees are exactly the \emph{increasing trees} rooted at $n{-}1$.

Let $\IncTrees(n-1)$ denote the set of increasing trees on vertex set $\cu{1,\ldots,n{-}1}$ rooted at $n{-}1$.
Then in $\Rnn$,
\begin{align}
    M_{1\cdots n}\Big|_{\Rnn}=\sum_{T\in\IncTrees(n-1)}\,\prod_{(u,v)\in E(T)}\ab{[uv]}.
\end{align}
This sum can be evaluated by the directed matrix-tree theorem, as we now show.

Consider the acyclic directed graph on the vertices $\cu{1,\ldots,n{-}1}$ with an arrow $i\to j$ for $i<j$, weighted by $w_{i\to j}=\ab{[ij]}=[ji]$.
Let $Q$ be the associated directed Laplacian matrix:
\begin{align}
    Q_{ij}=
    \begin{cases}
    [ij],& i\neq j,\ i<j,\\
    0,& i>j,\\
    \sum_{k=i+1}^{n-1}[ki],& i=j.
    \end{cases}
\end{align}
Then the directed matrix-tree theorem gives
\begin{align}
    \sum_{T\in\IncTrees(n-1)}\,\prod_{(u,v)\in E(T)}\ab{[uv]}
    =\det\pa{Q^{(n-1)}},
\end{align}
where $Q^{(n-1)}$ deletes row and column $n{-}1$.
Since $Q^{(n-1)}$ is upper triangular, its determinant factorizes and hence we recover~\eqref{eq:ProductFormula}:
\begin{align}
     M_{1\cdots n}\Big|_{\Rnn}=\det\pa{Q^{(n-1)}}
    =\prod_{i=1}^{n-2}\left(\sum_{j=i+1}^{n-1}[ji]\right).
\end{align}

\acknowledgments

We are grateful to Nima Arkani-Hamed, Filipe de Avila Belbute-Peres, Freddy Cachazo, Juan Maldacena, Mark Spradlin and Anastasia Volovich for useful discussions.
AL was supported in part by NSF grant AST-2307888, the NSF CAREER award PHY-2340457, and the Simons Foundation grant SFI-MPS-BH-00012593-09.
AG acknowledges the Roger Dashen membership at the IAS, and additional support from DOE grant DE-SC/0009988.
AS \& DS are supported in part by the Simons Collaboration for Celestial Holography.
DS is also supported by the STFC (UK) grant ST/X000664/1.
AS is supported by the Black Hole Initiative and NSF grant PHY–2207659 and is grateful for the hospitality of OpenAI.

\clearpage

\appendix
\onecolumngrid

\section{Review of MHV amplitudes and Cayley trees}\label{app:MHV}

To set notation, we first review the $n$-point MHV graviton amplitude~\cite{Berends:1988zp,Mason:2008jy,Feng:2012sy,Nguyen:2009jk,Hodges:2011wm,Miller:2024oza,Miller:2025wpq}.
It has two minus-helicity legs; without loss of generality, we take them to be $n{-}1$ and $n$.
With our normalization \eqref{eq:DeltaNormalization}, we write
\begin{align}
    \mathcal{M}_n^{\rm MHV}\pa{1^+,\ldots,(n{-}2)^+,(n{-}1)^-,n^-}=i\av{n{-}1\,n}^8\tilde{M}_n\,\delta^4\!\pa{\sum_{k=1}^np_k},
\end{align}
where $\tilde{M}_n$ is the stripped MHV amplitude.
We use this example to introduce key concepts used in the main body.

\paragraph{Cayley tree formula.}

Let $\mathsf{Trees}(S)$ denote the set of (unrooted) labeled trees whose vertex set is the finite set $S$.
As elements of $S$ will be physical particle labels, we refer to these trees as \textit{Cayley} trees, in contrast to Feynman trees.
Moreover, their union covers the complete graph of $S$ and in that sense they are \textit{spanning}.
For $\Gamma\in\mathsf{Trees}(S)$, we write $E(\Gamma)$ for its edge set.
Away from collinear points, the permutation-invariant tree formula for the stripped MHV amplitude is \cite{Nguyen:2009jk}\footnote{Only $S_{n-2}$ symmetry is manifest.}
\begin{align}
    \label{eq:MHVtree}
    \tilde{M}_n=\frac{1}{\av{n{-}1\,n}^2t_1^2\cdots t_{n-2}^2}\sum_{T\in\mathsf{Trees}(S)}\br{\prod_{(a,b)\in E(T)}t_a t_b\frac{[ab]}{\av{ab}}},
\end{align}
where $t_a=\av{an}\av{a\,n{-}1}$ and $S=\{1,\ldots,n-2\}$.
Applying the matrix-tree theorem to \eqref{eq:MHVtree} yields Hodges' determinant formula~\cite{Hodges:2011wm}.

\paragraph{Regularized Cayley factors.}

The ratios $[ij]/\av{ij}$ may be rewritten as $[ij]^2/p_{ij}^2$.
In a collinear regime where $\av{ij}\to0$, the correct distributional prescription is obtained by keeping the Feynman $i\epsilon$,
\begin{align}
    \frac{[ij]}{\av{ij}}
    \quad\longrightarrow\quad
    \frac{[ij]^2}{p_{ij}^2+i\epsilon}=\frac{[ij]}{z_{ij}+i\epsilon\sg_{ij}},\qquad
    \sg_{ij}:=\sg\pa{\br{ij}},
\end{align}
and we will repeatedly encounter the associated \emph{Cayley tree factor} 
\begin{align}
    \label{eq:CayleyFactor}
    \mathrm{T}_S:=\sum_{\Gamma\in\mathsf{Trees}(S)}\prod_{(i,j)\in E(\Gamma)}\frac{[ij]^2}{p_{ij}^2+i\epsilon}.
\end{align}
These objects play the same role in gravity that Parke--Taylor factors play in Yang--Mills theory.

\section{Derivation from Berends--Giele recursion}
\label{app:Derivation}

This appendix derives the recursion~\eqref{eq:Recursion} from the unordered Berends--Giele recursion~\cite{Berends:1987me} for gravity and provides the Cayley tree identities used throughout.

\paragraph{Unordered Berends--Giele recursion for gravity.}

Let $\mathcal{F}_S$ denote the planar gravitational form factor (current) with external set $S=\cu{1,\ldots,n{-}1}$ on-shell and one additional leg $n$ off-shell, in the self-dual sector.
The unordered Berends--Giele recursion relation takes the form
\begin{align}
    \label{eq:BG}
    \mathcal{F}_{\{i\}}=1,\qquad 
    \mathcal{F}_S=\frac{1}{P_S^2+i\epsilon}\sum_{\substack{A\sqcup B=S\\A,B\neq\emptyset}} [A,B]^2\,\mathcal{F}_A\,\mathcal{F}_B,
\end{align}
where $P_S=\sum_{i\in S}p_i$ and $[A,B]=[\tlam_A\tlam_B]$ with $\tlam_A=\sum_{i\in A}\tlam_i$.
The sum is over unordered bipartitions.

The single-minus amplitude is obtained by LSZ reduction on the remaining leg $n$,
\begin{align}
    \mathcal{M}_{1\cdots n}=\lim_{p_n^2\to 0}-ip_n^2\,\mathcal{F}_{\{1,\ldots,n-1\}}\,\delta^4\!\pa{\sum_{i=1}^n p_i},\qquad
    p_n=-\sum_{i=1}^{n-1}p_i.
\end{align}

\paragraph{Momentum identity.}

For any nonempty set $S$,
\begin{align}
    \label{eq:PS2Pairs}
    P_S^2=\sum_{\substack{i<j\\ i,j\in S}}\av{ij}[ij].
\end{align}
Let $T$ be a labeled spanning tree on vertex set $S$, and for an edge $e=(u,v)\in E(T)$, write $A_e\sqcup B_e=S$ for the cut induced by removing $e$, with $u\in A_e$ and $v\in B_e$.
Then the following identity holds for every tree $T$:
\begin{align}
    \label{eq:CayleyIdentity}
    P_S^2=\sum_{e=(u,v)\in E(T)}\av{uv}[A_e,B_e].
\end{align}

To show this, fix the frame~\eqref{eq:Conventions}, so $\av{ij}=z_{ij}$ and $p_{ij}^2=z_{ij}[ij]$.
In this frame, $[A_e,B_e]=\sum_{i\in A_e}\sum_{j\in B_e}[ij]$ and write the RHS of~\eqref{eq:CayleyIdentity} as a sum over pairs $(i,j)$:
\begin{align}
    \sum_{e=(u,v)\in E(T)}z_{uv}[A_e,B_e]
    =\sum_{i<j\in S}[ij]\sum_{\substack{e=(u,v)\in E(T)\\ \text{$e$ separates $i$ and $j$}}}z_{uv}.
\end{align}
In a tree, the edges separating $i$ and $j$ are precisely the edges along the unique path from $i$ to $j$.
Along that path, the sum of differences $z_{uv}=z_u-z_v$ telescopes to $z_i-z_j=\langle ij\rangle$ (up to a sign fixed by the path orientation, which matches the separation convention).
Thus, the coefficient of $[ij]$ is $z_{ij}$ and the sum equals~\eqref{eq:PS2Pairs}.

\paragraph{Distributional identity.}

The Cayley tree factor \eqref{eq:CayleyFactor} is, outside the half-collinear regime, a rational function of the kinematics.
In the half-collinear regime, it develops distributional support governed by the identities in App.~A of \cite{Guevara:2026qzd}.
Using this in combination with the tree identity~\eqref{eq:CayleyIdentity}, one derives the following analogue of the Parke--Taylor identity of \cite{Guevara:2026qzd}:
\begin{align}
    \label{eq:TreeIdentity}
    \TT_S-\delta_S\,V_{S}=\frac{1}{P_S^2+i\epsilon}\sum_{\substack{A\sqcup B=S\\A,B\neq\emptyset}}[A,B]^2\,\TT_A\,\TT_B,
\end{align}
where we recall the definition \eqref{eq:DeltaS} of the fully collinear object $\delta_S$.
Here, $V_{\tlam_S}$ is the retarded vertex obtained by localization $\TT_S$ onto the fully collinear kinematics, which can be expressed as
\begin{align}
    V_{S}=\sum_{T\in\Trees(S)}\ \prod_{e=(u,v)\in E(T)}\ab{[uv]}\,\Theta\pa{-\,\frac{[A_e,B_e]}{[uv]}}.
\end{align}
The advanced vertex $\bar{V}_{S}$ is obtained by reversing the sign in the $\Theta$-argument.
The identity \eqref{eq:TreeIdentity} is the key input for solving the BG recursion by ``replacing one $V$ by $\TT$'' and simultaneously controlling the half-collinear contact terms.

\paragraph{General form factor and on-shell recursion.}

The solution of~\eqref{eq:BG} can be written as a sum over set partitions of $S$.
Let $S=S_1\sqcup\cdots\sqcup S_A$ be a partition into $A\ge 1$ nonempty blocks, and let $\tlam_{S_a}=\sum_{i\in S_a}\tlam_i$ as usual.
Then
\begin{align}
    \label{eq:FormFactorSolution}
    \mathcal{F}_{S}=\sum_{\substack{S=S_1\sqcup\cdots\sqcup S_A\\A\ge 1}}\TT_{S_1\cdots S_A}\,\prod_{a=1}^{A}\Big(\bar{M}_{S_a}\,\delta_{S_a}\Big),
\end{align}
where $\TT_{S_1\cdots S_A}$ is the Cayley factor built from the block momenta:
\begin{align}
    \TT_{S_1\cdots S_A}:=\sum_{T\in\Trees(A)}\ \prod_{(a,b)\in E(T)}\frac{\br{S_aS_b}}{z_{ab}+i\epsilon\sg_{ab}},
    \qquad
    \sg_{ab}:=\sg\pa{\br{S_aS_b}}.
\end{align}
Inserting~\eqref{eq:FormFactorSolution} into~\eqref{eq:BG} and using~\eqref{eq:TreeIdentity} yields the preamplitude recursion~\eqref{eq:Preamplitudes}.

Finally, LSZ reduction on the remaining leg $n$ evaluates the on-shell limit of $\TT_{S_1\cdots S_A}$.
Using the master identity from App.~A of \cite{Guevara:2026qzd} together with~\eqref{eq:CayleyIdentity}, one finds, with $\hTT=V-\bar{V}$ as in~\eqref{eq:HatT} and $\delta_{S,n}$ the half-collinear support on the union $S\cup\cu{n}$,
\begin{align}
    \label{eq:LSZHatT}
    \lim_{p_n^2\to 0}p_n^2\,\TT_{S_1\cdots S_A}\,\delta^4\!\pa{\sum_{j=1}^n p_j}
    =\hTT_{S_1\cdots S_A}\,\delta_{S_1,\ldots,S_A, n}\,\delta^2\!\pa{\sum_{j=1}^n\tlam_j}.
\end{align}
Stripping off the universal support then gives the on-shell recursion~\eqref{eq:Recursion}. The explicit solution is
\begin{align}
    \label{eq:Full}
    M_{1\cdots n}=\sum_{\tau\in\mathrm{FT}_n(r)}\br{-\widehat{\mathrm T}_{\tlam_{S_1(r)}\,\cdots\,\tlam_{S_{d(r)}(r)}}\,\prod_{v\in\mathrm{Int}(\tau)\setminus\{r\}}\pa{-V_{\tlam_{S_1(v)}\,\cdots\,\tlam_{S_{d(v)}(v)}}}},
\end{align}
where $\mathrm{FT}_n(r)$ denotes the (unordered) Feynman trees of leaves $\cu{1,\ldots,n{-}1}$ with a root labeled by $r$, with $d(r)\geq2$ edges connected to $r$ and $d(v)\geq3$ edges connected to vertex $v$.
For each internal vertex ($r$ or $v$), the set of 2D momenta incoming onto it are labeled $\tlam_{S_j(v)}$.
A similar formula exists for gluons, summing instead over (color-)ordered trees.

\section{\texorpdfstring{$\mathcal{L}w_{1+\infty}$}{Lw(1+∞)} recursion}
\label{app:ward-recursion-proof}

\paragraph{Setup and statement on a fixed chamber.}

Fix $n\ge 3$ and a connected component (``chamber'')
\begin{align}
    \Rnn^\pi\subset\mathcal{D}_{n,n-1}\setminus\Sigma_n,
\end{align}
where $\Sigma_n$ is the singular locus (coincidence walls and possible additional bulk walls).
By analyticity within $\Rnn^\pi$, the signs of all relevant brackets are fixed on $\Rnn^\pi$, so absolute values are locally linear and the shift maps below do not cross walls.

For any inserted plus-helicity leg $s\in\cu{1,\dots,n{-}2}$, the integrated $\mathcal{L}w_{1+\infty}$ Ward identity on $\Rnn^\pi$ takes the shift form \eqref{eq:Ward}, which we quote here for readability:
\begin{align}
    \label{eq:WardShift}
    M_{1\cdots n}=\frac{1}{2}\sum_{\substack{i=1\\ i\neq s}}^{n}|[si]|\,\Big(M_{\widehat{s}}\Big)^{(i)},\qquad
    \Big(M_{\widehat{s}}\Big)^{(i)}:=M_{\widehat{s}}\Big|_{\tlam_i\to \tlam_i+\tlam_s},
\end{align}
where $M_{\widehat{s}}$ is the stripped amplitude with leg $s$ removed.
We will choose a momentum-conserving representative such that $M_{\widehat{s}}$ has no explicit dependence on $\tlam_n$, hence $\Big(M_{\widehat{s}}\Big)^{(n)}=M_{\widehat{s}}$.
We define the decay ansatz $M_{1\cdots n}^{\rm dec}$ by
\begin{align}
    \label{eq:DecayAnsatz}
    M_{1\cdots n}^{\rm dec}=\prod_{a=1}^{n-2}S_a,\qquad\text{where}\qquad
    S_a=\sum_{j:[ja]>0}[ja],\qquad
    a\in\cu{1,\dots,n{-}2}.
\end{align}
The goal is to prove that within each chamber $\Rnn^\pi$, $M_{1\cdots n}\big|_{\Rnn^\pi}=M_{1\cdots n}^{\rm dec}\big|_{\Rnn^\pi}$, so that the recursion \eqref{eq:WardShift} generates $M^{\rm dec}$ with seed $M_{123}=|[12]|$.
The argument follows from induction on $n$ starting from $n=3$.
Recall that to simplify the algebra, we work in a representative ordered chamber where $\pi=(1,\ldots,n{-}2)$, \emph{i.e.},
\begin{align}
    \label{eq:TrivialPermutation}
    \tilde{z}_1<\tilde{z}_2<\cdots<\tilde{z}_{n-2}<\tilde{z}_n<\tilde{z}_{n-1},
\end{align}
and then use label covariance/permutation invariance to cover the other chambers in $\mathcal{D}_{n,n-1}$ at the end.

\paragraph{Reduction of \texorpdfstring{$S_a$}{S(a)} inside an ordered chamber.}

Within this chamber, one has $[aj]>0$ for $j<a$, $[aj]<0$ for $a<j\le n{-}1$, and $[an]>0$ for $a\le n{-}2$.
Hence we will write $S_a$ as
\begin{align}
    \label{eq:Sans}
    S_a
    =\sum_{j=a+1}^{n-1}|[aj]|=|[an]|+\sum_{j=1}^{a-1}|[aj]| .
\end{align}
Thus, in the ordered chamber~\eqref{eq:TrivialPermutation}, the decay ansatz for the stripped amplitudes with graviton $s$ removed is
\begin{align}
    \label{eq:Target}
    M_{\widehat{s}}^{\rm dec}=\pa{\prod_{a<s}\pa{S_a-|[as]|}}\pa{\prod_{a>s}S_a},
\end{align}
where leg $i$ has not yet been shifted.

\subsection{Telescoping verification of the shift Ward identity}

We now use \eqref{eq:Target} to prove that $M_{1\cdots n}^{\rm dec}$ in \eqref{eq:DecayAnsatz} satisfies the shifted Ward identity \eqref{eq:WardShift} for any $s\in\cu{1,\dots,n{-}2}$.
The proof consists of splitting the sum in \eqref{eq:WardShift} into two terms, a right part and a left part.

\paragraph{Right part: \texorpdfstring{$i=s+1,\dots,n{-}1$}{i=s+1,...,n-1}.}

Define tail sums
\begin{align}
    R_i:=\sum_{b=i+1}^{n-1}|[sb]|,\qquad
    \forall\,i\in\cu{s,\dots,n{-}1},\qquad
    R_{n-1}=0.
\end{align}
Then for $i\geq s+1$, we have $|[si]|=R_{i-1}-R_i$.
When $i\in\cu{s{+}1,\dots,n{-}2}$, the shift $\tlam_i\to\tlam_i+\tlam_s$ acts linearly on $|[ij]|$ inside the ordered chamber, because we assume $\omega_s/\omega_i\ll 1$.
Applying the shift to \eqref{eq:Target} gives
\begin{align}
    \Big(M^{\rm dec}_{\widehat{s}}\Big)^{(i)}=\pa{\prod_{a<s}S_a}\pa{\prod_{s<a<i}\pa{S_a-|[sa]|}}(S_i+R_i)\pa{\prod_{a>i}S_a},\qquad
    \forall\,i\in\cu{s{+}1,\dots,n{-}2},
\end{align}
while in the special case $i=n{-}1$,
\begin{align}
    \Big(M^{\rm dec}_{\widehat{s}}\Big)^{(n-1)}=\pa{\prod_{a<s}S_a}\pa{\prod_{s<a\le n-2}\pa{S_a-|[sa]|}}.
\end{align}
Now define
\begin{align}
    T_i:=R_i\pa{\prod_{a<s}S_a}\pa{\prod_{s<a\le i}(S_a-|[sa]|)}\pa{\prod_{a>i}S_a},\qquad
    \forall\,i\in\cu{s,\dots,n{-}1}.
\end{align}
A direct algebraic subtraction gives, for $i\in\cu{s{+}1,\dots,n{-}2}$,
\begin{align}
    |[si]|\,\Big(M_{\widehat{s}}\Big)^{(i)}=T_{i-1}-T_i,
\end{align}
and for $i=n{-}1$, the same holds because $T_{n-1}=0$ and $R_{n-2}=|[s\,n{-}1]|$.
Therefore, the right sum telescopes:
\begin{align}
    \label{eq:RightTelescope}
    \sum_{i=s+1}^{n-1}|[si]|\,\Big(M_{\widehat{s}}\Big)^{(i)}=\sum_{i=s+1}^{n-1}(T_{i-1}-T_i)
    =T_s
    =R_s\prod_{a\neq s}S_a
    =S_s\prod_{a\neq s}S_a
    =M_{1\cdots n}^{\rm dec}.
\end{align}

\paragraph{Left part: \texorpdfstring{$i=1,\dots,s{-}1$}{i=1,...,s-1} plus the \texorpdfstring{$i=n$}{i=n} term.}

Similarly, we define head sums, for $i\in\cu{0,\dots,s{-}1}$,
\begin{align}
    L_i:=\sum_{b=1}^{i}|[bs]|,\qquad
    L_0:=0,
\end{align}
so that $|[si]|=L_i-L_{i-1}$ for $1\le i\le s{-}1$.
From \eqref{eq:Sans} with $a=s$,
\begin{align}
    \label{eq:sn}
    |[sn]|=S_s-L_{s-1}.
\end{align}
For $i\in\cu{1,\dots,s{-}1}$, one can explicitly evaluate the shifted product (the only nontrivial change is in the factor $(S_i-|[is]|)$):
\begin{align}
    \Big(M_{\widehat{s}}\Big)^{(i)}=\pa{\prod_{a<i}S_a}\pa{S_i+|[sn]|+L_{i-1}}\pa{\prod_{i<a<s}\pa{S_a-|[as]|}}\pa{\prod_{a>s}S_a},\qquad
    1\le i\le s{-}1.
\end{align}
Now define
\begin{align}
    U_i:=\pa{|[sn]|+L_i}\pa{\prod_{a\le i}S_a}\pa{\prod_{i<a<s}\pa{S_a-|[as]|}}\pa{\prod_{a>s}S_a},\qquad
    i\in\cu{0,\dots,s{-}1}.
\end{align}
Using $L_i-L_{i-1}=|[si]|$ and the fact that $U_{i-1}$ contains $(S_i-|[is]|)$ in the second product while $U_i$ contains $S_i$ in the first product, one obtains the telescoping identity
\begin{align}
    \label{eq:Telescoping}
    |[si]|\,\Big(M_{\widehat{s}}\Big)^{(i)}=U_i-U_{i-1}\qquad
    i\in\cu{1,\dots,s{-}1},\qquad
    |[sn]|\,\Big(M_{\widehat{s}}\Big)^{(n)}=|[sn]|M_{\widehat{s}}=U_0,
\end{align}
where we used \eqref{eq:DecayAnsatz} for the $i=n$ term. Summing \eqref{eq:Telescoping} yields
\begin{align}
    \label{eq:LeftTelescope}
    \sum_{i=1}^{s-1}|[si]|\,\Big(M_{\widehat{s}}\Big)^{(i)}+|[sn]|\,M_{\widehat{s}}=\sum_{i=1}^{s-1}\pa{U_i-U_{i-1}}+U_0
    =U_{s-1}.
\end{align}
Finally, by \eqref{eq:sn}, one has $|[sn]|+L_{s-1}=S_s$, so that
\begin{align}
    \label{eq:LeftTelescopeBis}
    U_{s-1}=(|[sn]|+L_{s-1})\prod_{a\neq s}S_a
    =S_s\prod_{a\neq s}S_a
    =M.
\end{align}

Finally, combining \eqref{eq:RightTelescope} with \eqref{eq:LeftTelescope}--\eqref{eq:LeftTelescopeBis} gives
\begin{align}
    \sum_{\substack{i=1\\ i\neq s}}^{n}|[si]|\,\Big(M_{\widehat{s}}\Big)^{(i)}=\underbrace{\sum_{i=s+1}^{n-1}|[si]|\,\Big(M_{\widehat{s}}\Big)^{(i)}}_{=\,M}+\underbrace{\Big(\sum_{i=1}^{s-1}|[si]|\,\Big(M_{\widehat{s}}\Big)^{(i)}+|[sn]|M_{\widehat{s}}\Big)}_{=\,M}
    =2M.
\end{align}
Multiplying by the $\frac{1}{2}$ prefactor in \eqref{eq:WardShift} shows that $M_{1\cdots n}^{\rm dec}=\prod_{a=1}^{n-2}S_a$ satisfies the shifted Ward identity in the ordered chamber.

\bibliographystyle{utphys}
\bibliography{GR.bib}

\end{document}